\documentstyle[twocolumn,aps,epsbox]{revtex}       % documentstyle set

\def\dag{\dagger}
\def\mathrm#1{{\rm #1}}

\def\diff#1#2{\dfrac{d #1}{d #2}}

\def\i{\mathop{\mathrm{i}}\nolimits}

\def\th{\mathop{\mathrm{th}}\nolimits}

\def\dis{\displaystyle}
\def\dfrac#1#2{{\frac{\dis #1}{\dis #2}}}
\def\mib#1{{\mbox{\boldmath $#1$}}}

\begin{document}
\draft

%----- Title ------
\title
{Theory of Unconventional Spin Density Wave:
A Possible Mechanism of the Micromagnetism in U-based Heavy Fermion Compounds}
\author
{Hiroaki Ikeda
\footnote{present address: Department of Physics, Kyoto University, Kyoto
606-8502, Japan}
and
Yoji Ohashi$^1$}
\address
{
Joint Research Center for Atom Technology (JRCAT),
National Institute for Advanced Interdisciplinary Research (NAIR),
1-1-4 Higashi, Tsukuba, Ibaraki 305, Japan\\
$^1$Institute of Physics, University of Tsukuba, Ibaraki 305, Japan
}

\maketitle

\begin{abstract}
We propose a novel spin density wave (SDW) state as a possible mechanism of the
anomalous antiferromagnetism, so-called the micromagnetism, in URu$_2$Si$_2$
below 17.5[K]. In this new SDW, the electron-hole pair amplitude changes its
sign in the momentum space as in the case of the unconventional
superconductivity. It is shown that this state can be realized in an extended
Hubbard model within the mean field theory. We also examine some characteristic
properties of this SDW to compare with the experimental results. All these
properties well explain the unsolved problem of the micromagnetism.
\end{abstract}


%----- main -------
%\narrowtext

%% Introduction %%
{\it Introduction}---
Electronic states in the U-based heavy fermion compounds UPt$_3$ and
URu$_2$Si$_2$ have recently attracted much attentions because of their curious
physical properties. The antiferromagnetism is one of them:
(1) Magnitude of the induced staggered magnetic moment is extremely small;
It is about 0.02$\mu_\mathrm{B}$ for UPt$_3$
($T_\mathrm{N}=5$[K])~\cite{rf:Aeppli,rf:Isaacs1} while 0.04$\mu_\mathrm{B}$
for URu$_2$Si$_2$ ($T_\mathrm{N}=17.5$[K])~\cite{rf:Broholm,rf:Isaacs2}.
They are about 1\% of the values expected from the magnetic susceptibility
at high temperatures~\cite{rf:Visser,rf:Palstra}.
(2) The phase transition can be observed by the neutron
scattering~\cite{rf:Aeppli,rf:Broholm} and the magnetic X-ray
diffraction~\cite{rf:Isaacs1,rf:Isaacs2} in both the materials;
in the case of UPt$_3$, however, it is not detected by other probes,
NMR-$T_1^{-1}$~\cite{rf:Tou}, the susceptibility, and the specific
heat~\cite{rf:Visser}. On the other hand,a clear jump in the specific heat at
$T_\mathrm{N}$ and rapid decreases in NMR-$T_1^{-1}$ and in the susceptibility
below $T_\mathrm{N}$ can be observed in the case of URu$_2$Si$_2$.
This indicates that, at least for URu$_2$Si$_2$, a magnetic phase transition
really occurs at $T_\mathrm{N}$.
\par
Although various mechanisms have been proposed in order to explain this
{\it micromagnetism}~\cite{rf:Santini,rf:Barzykin,rf:Sikkema}, this problem
still remains to be settled. Since the antiferromagnetism continues to exist
even in the unconventional superconducting phase, this problem is also related
to the mechanism of the superconductivity in these materials.
\par
Motivated by this situation, we study the micromagnetism and propose a new
mechanism in this paper. We mainly focus on URu$_2$Si$_2$, because its phase
transition is well-confirmed compared with UPt$_3$ as noted in the above.
Namely, we study {\it how the tiny moment and clear anomalies in some physical
quantities can coexist without any inconsistency}.
\par
%% Outline %%
{\it Unconventional SDW}---
First of all, we explain the outline of our idea.
(1) In the SDW, the ordered magnetic moment $M_Q$ is given by
\begin{equation}
M_Q=\sum_k\Psi_k^Q,
\label{intro}
\end{equation}
where $\Psi_k^Q\equiv\sum_{\sigma}\sigma\left< c^\dag_{k\sigma}c_{k+Q\sigma} \right>$
is the electron-hole pair amplitude.
($Q$ represents a characteristic wave vector of the SDW.)
In the ordinary simple SDW, Eq. (\ref{intro}) is finite, because $\Psi_k^Q$ is
a positive constant and is independent of $k$.
(2) In this paper, we notice the $k$-dependence of $\Psi_k^Q$, and propose an
{\it unconventional SDW} where $\Psi_k^Q$ changes its sign depending on $k$;
for example, the $d$-wave symmetry $\Psi_k^Q \propto \cos k_x - \cos k_y$ as
in the case of the $d$-wave superconductivity. In this case, Eq. (\ref{intro})
gives $M_Q=0$, because $\Psi_k^Q$ is canceled in the $k$-summation due to its
sign-change. (3) Presence of $\Psi_k^Q$ decreases the excitation spectrum below
the energy gap, even if $\Psi_k^Q$ is anisotropic and $M_Q$ is absent.
Then, this energy gap should induce {\it a large anomaly} in, for example,
the specific heat. Thus, ``absence of the ordered magnetic moment and presences
of large anomalies in some thermodynamic quantities'' can be realized in the
{\it unconventional SDW}.
\par
This is our basic idea for the micromagnetism in URu$_2$Si$_2$. In the next
section, we present a concrete example of this state based on a simple model.
\par
%% Hamiltonian %%
{\it Formulation}---
Since the micromagnetism occurs after the heavy fermion was
formed\cite{rf:Palstra,rf:Kimura,rf:Maple}, we examine the quasiparticle state
described by the Hamiltonian
\begin{eqnarray}
\label{eq:H}
&H& = -t\sum_{\left<ij\right>\sigma}
\left(c^\dag_{i\sigma}c_{j\sigma}+\mathrm{h.c.}\right)
    +U\sum_{i}n_{i\uparrow}n_{i\downarrow} \nonumber \\
&+&V\sum_{\left<ij\right>}\sum_{\sigma\sigma'}n_{i\sigma}n_{j\sigma'}
    +J\sum_{\left<ij\right>}\sum_{\sigma\sigma'}
       c^\dag_{i\sigma}c^\dag_{j\sigma'}c_{i\sigma'}c_{j\sigma},
\end{eqnarray}
where $c^\dag_{i\sigma}$ is the creation operator of a quasiparticle and
$n_{i\sigma}=c^\dag_{i\sigma}c_{i\sigma}$.
The band structure is controlled by the hopping term in Eq.~(\ref{eq:H}), in
which $\left<ij\right>$ represents the summation over nearest-neighbor pairs;
although URu$_2$Si$_2$ is a three dimensional material and should have
a complex band structure, we use a two-dimensional simple square lattice
in order to grasp the essence of the unconventional SDW. Furthermore, we put
the chemical potential $\mu$ equal to zero: The Fermi surface is then in the
perfect nesting with $\mib{Q}=(\pi,\pi)$. On the other hand, we take into
account three kinds of interactions; the on-site Coulomb repulsion $U$, the
nearest-neighbor direct interaction $V~(>0)$, and the exchange one $J~(>0)$.
\par
Within the mean field theory, the most possible ordered states which may be
realized in Eq.~(\ref{eq:H}) are the conventional SDW, the charge density wave
(CDW), and three kinds of novel SDW's, all of that characterized by the nesting
vector, $\mib{Q}=(\pi,\pi)$~\cite{rf:Ozaki}. The mean field Hamiltonian for Eq.
(\ref{eq:H}) is then given by
\begin{eqnarray}
\label{eq:Hq}
H &=& \sum_{k\sigma} \xi_k c^\dag_{k\sigma} c_{k\sigma}
   +\dfrac{1}{2}\sum_k\left[\Delta^\mathrm{CDW}\rho_{kk+Q}+\mathrm{h.c.}\right]
  \nonumber \\
  &+& \dfrac{1}{2}\sum_{k\alpha}
   \left[\Delta_{k\alpha}^\mathrm{SDW}{\sigma_{kk+Q}^{z}}+\mathrm{h.c.}\right],
\end{eqnarray}
where $\xi_k=-2t(\cos k_x + \cos k_y)$,
$\rho_{kk+Q}=c^\dag_{k\uparrow}c_{k+Q\uparrow}+c^\dag_{k\downarrow}c_{k+Q\downarrow}$ and $\sigma_{kk+Q}^z=c^\dag_{k\uparrow}c_{k+Q\uparrow}-c^\dag_{k\downarrow}c_{k+Q\downarrow}$.
(We have chosen the direction of the SDW order parameters being parallel to the
$z$-axis without loss of generality.) The CDW order parameter,
$\Delta^\mathrm{CDW}$, and the four SDW ones specified by the channel index
$\alpha$, $\Delta^\mathrm{SDW}_{k\alpha}$, are respectively given by
\begin{equation}
\label{eq:GAP}
\left\{
\begin{array}{l}
\dis
\Delta^\mathrm{CDW}=\dfrac{8V-U-4J}{2}\sum_{k'\sigma}
      \left<\rho_{k'+Q k'}\right>,
\nonumber \\
\dis
\Delta^\mathrm{SDW}_{k\alpha}
=\dfrac{I_\alpha}{2}\phi_k^\alpha\sum_{k'\sigma}\phi_{k'}^\alpha
      \left<\sigma^z_{k'+Q k'}\right>.
\end{array}
\right.
\end{equation}
In Eq. (\ref{eq:GAP}), $\phi_k^\alpha$ and $I_\alpha$ respectively represent
the basis function which determines the symmetry of the order parameter and
the corresponding pairing interaction. For the conventional ``$s$-wave'' SDW,
$\phi_k^s=1$ and $I_s=U-4J$. On the other hand, the nearest-neighbor direct
interaction in Eq. (\ref{eq:H}) gives three possible ``unconventional'' SDW's;
(1) $d$-wave: $\phi_k^d=\cos k_x - \cos k_y$. (2) Extended $s$-wave:
$\phi_k^{ex}=\cos k_x + \cos k_y$. (3) $p$-wave:
$\phi_k^p=\sqrt{2}\sin k_{x,y}$. All the three states have the same pairing
interaction, $I_\alpha=V$.
\par
\begin{figure}
\begin{center}
\psbox[height=6cm]{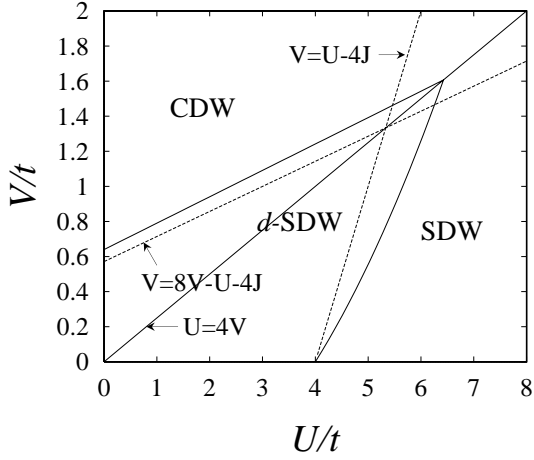}
\end{center}
\caption{$U-V$ phase diagram at $J=t$. All the states in the figure are
characterized by the nesting vector $\mib{Q}=(\pi,\pi)$.}
\label{fig:1}
\end{figure}

All the states obey the same form of the self-consistency equation,
\begin{equation}
\label{eq:Gapeq}
1=I_\alpha\sum_{k}\dfrac{\phi_k^{\alpha 2}}{2E_k}\th\dfrac{E_k}{2T}
~~~\left(E_k=\sqrt{\xi_k^2+|\Delta^\mathrm{SDW}_{k\alpha}|^2}\right).
\end{equation}
(For the CDW, we replace $I_\alpha\to 8V-U-4J$ and
$\Delta^\mathrm{SDW}_{k\alpha}\to \Delta^\mathrm{CDW}$) Then, neglecting the
anisotropy of the Fermi surface and comparing the pairing interactions only, we
immediately find that the unconventional SDW's are the most stable in the range
of $U-4J < V < (U+4J)/7$~\cite{rf:Ozaki}. Anisotropy of the Fermi surface
widens this region, and furthermore the $d$-wave SDW becomes the most favorable
among the unconventional SDW's, because the $d$-wave basis function,
$\phi_k^d=\cos k_x - \cos k_y$, has a large value at the corner of the Fermi
surface at which the density of states diverges. We demonstrate a typical
$U$-$V$ phase diagram at $J=t$ in Fig.~\ref{fig:1}. Clearly, there exists
a stable region of the $d$-wave SDW which is wider than that of the above
simple evaluation.
\par
At this stage, it is difficult to evaluate $U$, $V$ and $J$ for URu$_2$Si$_2$.
However, in real materials, at least the on-site $U$ should be larger than the
other nearest-neighbor interactions, $V$ and $J$. Figure~\ref{fig:1} shows that
the $d$-wave SDW can be realized even under this physical restriction,
$U> V, J$. Thus, the $d$-wave SDW can be considered as a possible state in real
systems.
\par
%% Physical Properties %%
{\it Physical properties of the $d$-wave SDW}---
Let us proceed to the detail analysis of some physical properties of the
$d$-wave SDW comparing with the conventional $s$-wave one. In what follows,
we simply write $\Delta^\mathrm{SDW}_{k\alpha=d,s}$ as $\Delta_k^{d,s}$.
\par
{\it (1) SDW order parameter:}
The $d$-wave SDW order parameter, $\Delta^d_k$ must be a purely imaginary
number contrary to the conventional SDW, because $\Delta_k^d$ satisfies
$(\Delta_k^d)^*=-\Delta_k^d$ due to $\phi_{k+Q}^d=-\phi_k^d$. (Note that one
cannot choose the phase of the order parameter freely in contrast to the case
of the superconductivity.) Because of this property, the $d$-wave SDW has
a finite spin current as pointed out by Ozaki\cite{rf:Ozaki}.
\par
\begin{figure}
\begin{center}
\psbox[height=6cm]{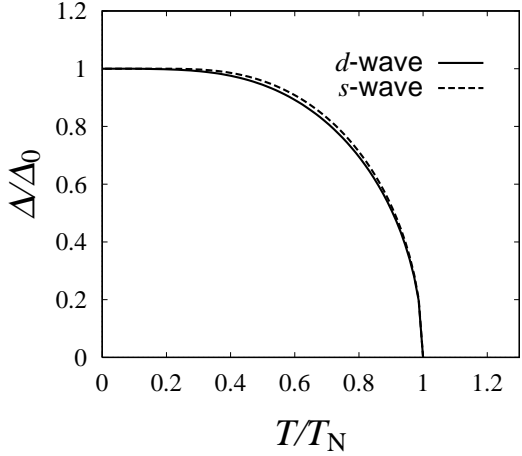}
\end{center}
\caption{Temperature dependence of the order parameter normalized by
$\Delta_0 \equiv \Delta(T=0)$ of each state. Relation between $\Delta$ in the
figure and the order parameter is $\Delta^d_k=\i\phi_k\Delta$ in the case of
the $d$-wave. We put $(U-4J)/t=V/t=2$, which gives $T_\mathrm{N}\simeq 0.4t$
($0.2t$) for the $d$-wave ($s$-wave). We also use the same parameter set
in Fig.~3.}
\label{fig:2}
\end{figure}

Figure~\ref{fig:2} shows the temperature dependence of the SDW order parameter.
It is found that there is no essential difference in between the $d$-wave SDW
and the $s$-wave one. In the $d$-wave case, the ratio of the order parameter to
$T_\mathrm{N}$ is
\begin{equation}
{2\Delta_\mathrm{max} \over T_\mathrm{N}} \simeq 4.8
~~~\left(\Delta_\mathrm{max} \equiv \max(|\Delta^d_k|) \right).
\end{equation}
Thus this state has a large excitation gap with the order of $T_\mathrm{N}$.
Such a gap is actually observed by the inelastic neutron
scattering~\cite{rf:Broholm,rf:Isaacs2}. (Strictly speaking, since the $d$-wave
order parameter has nodes, the density of states is not completely absent even
below $\Delta_\mathrm{max}$.)
\par
{\it (2) Thermodynamic properties:}
We examine how the presence of the excitation gap affects the thermodynamic
property below $T_\mathrm{N}$.  Here, we show two examples:
\par
\begin{figure}
\begin{center}
\psbox[height=6cm]{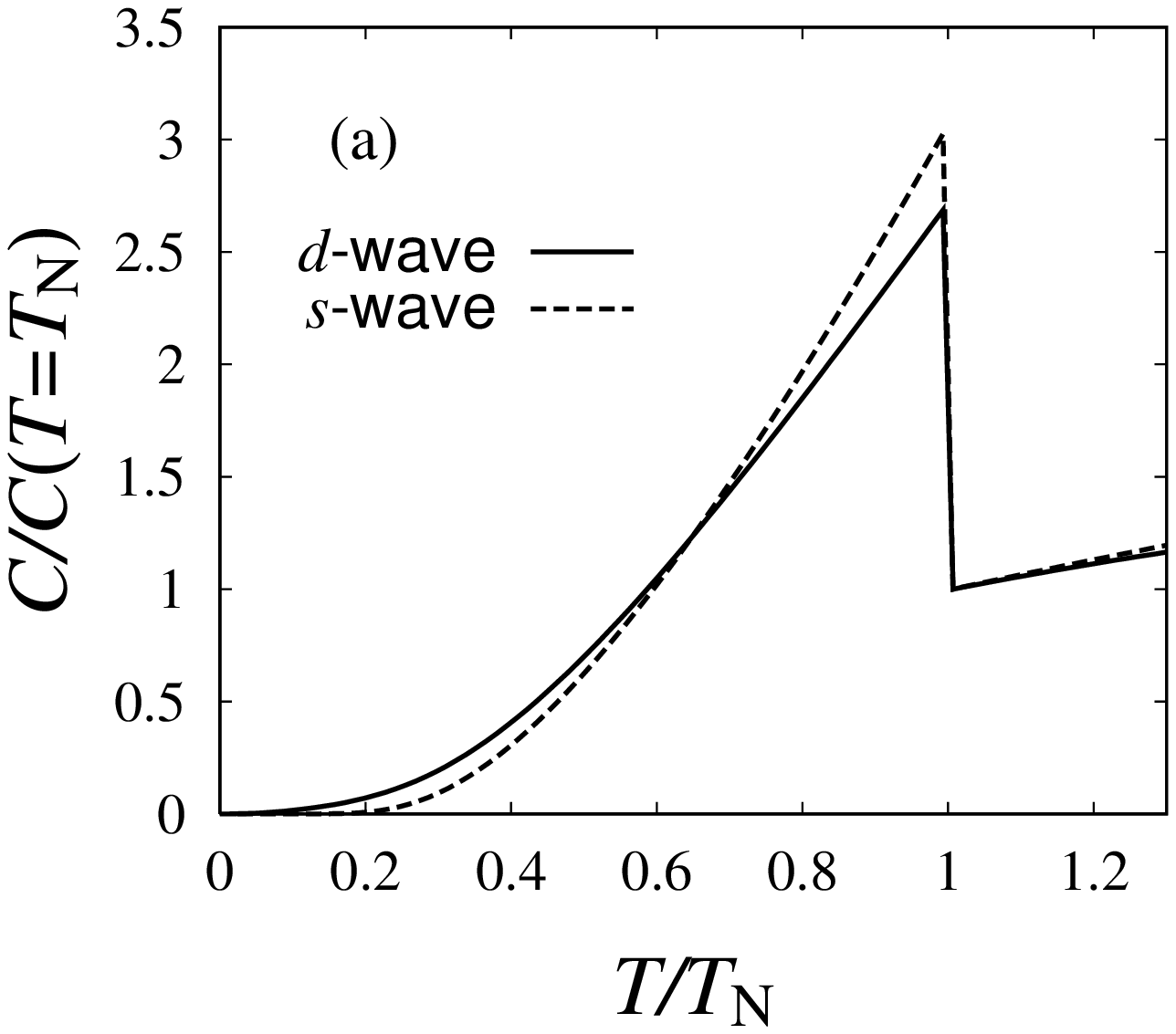}
\psbox[height=6cm]{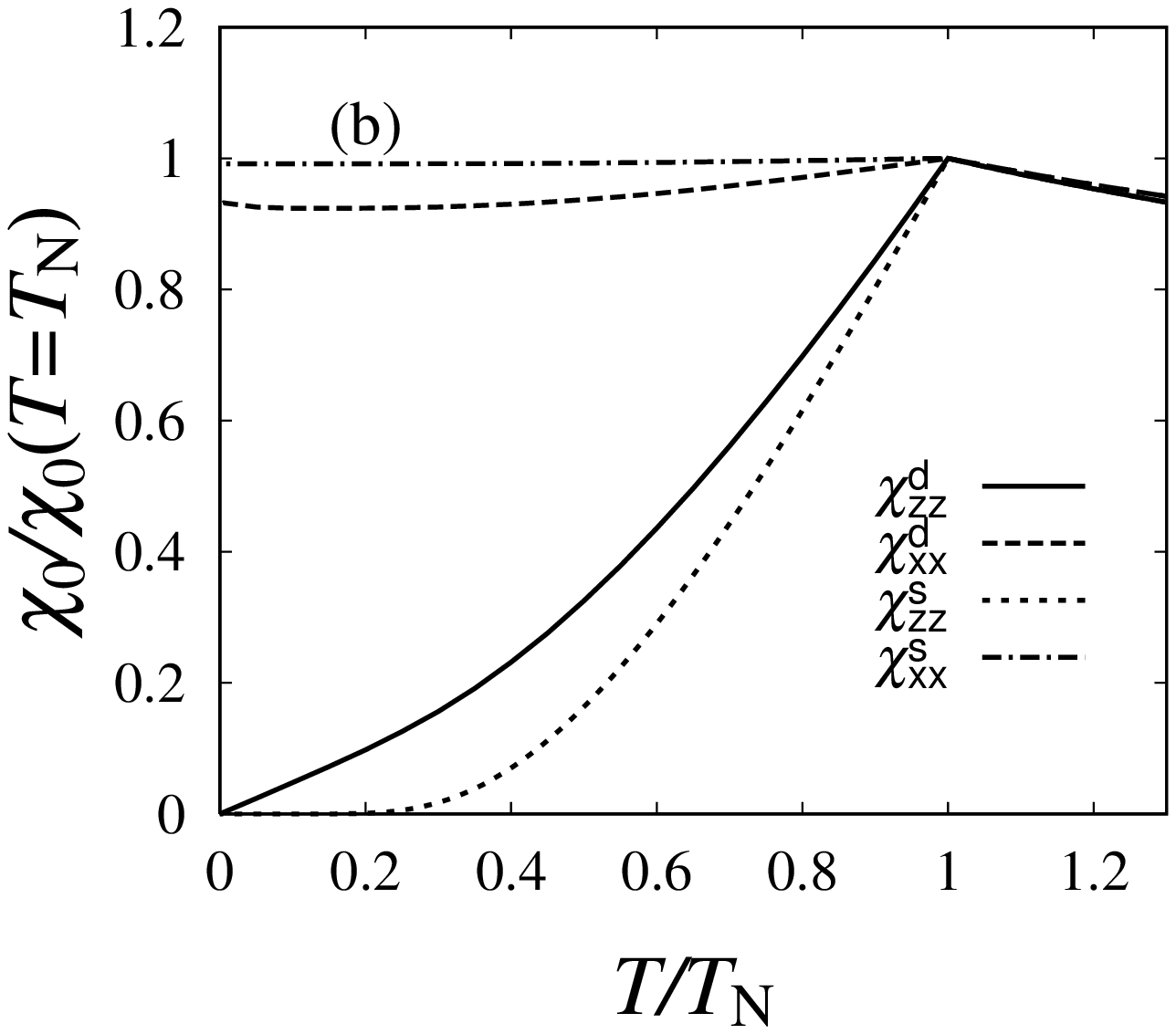}
\psbox[height=6cm]{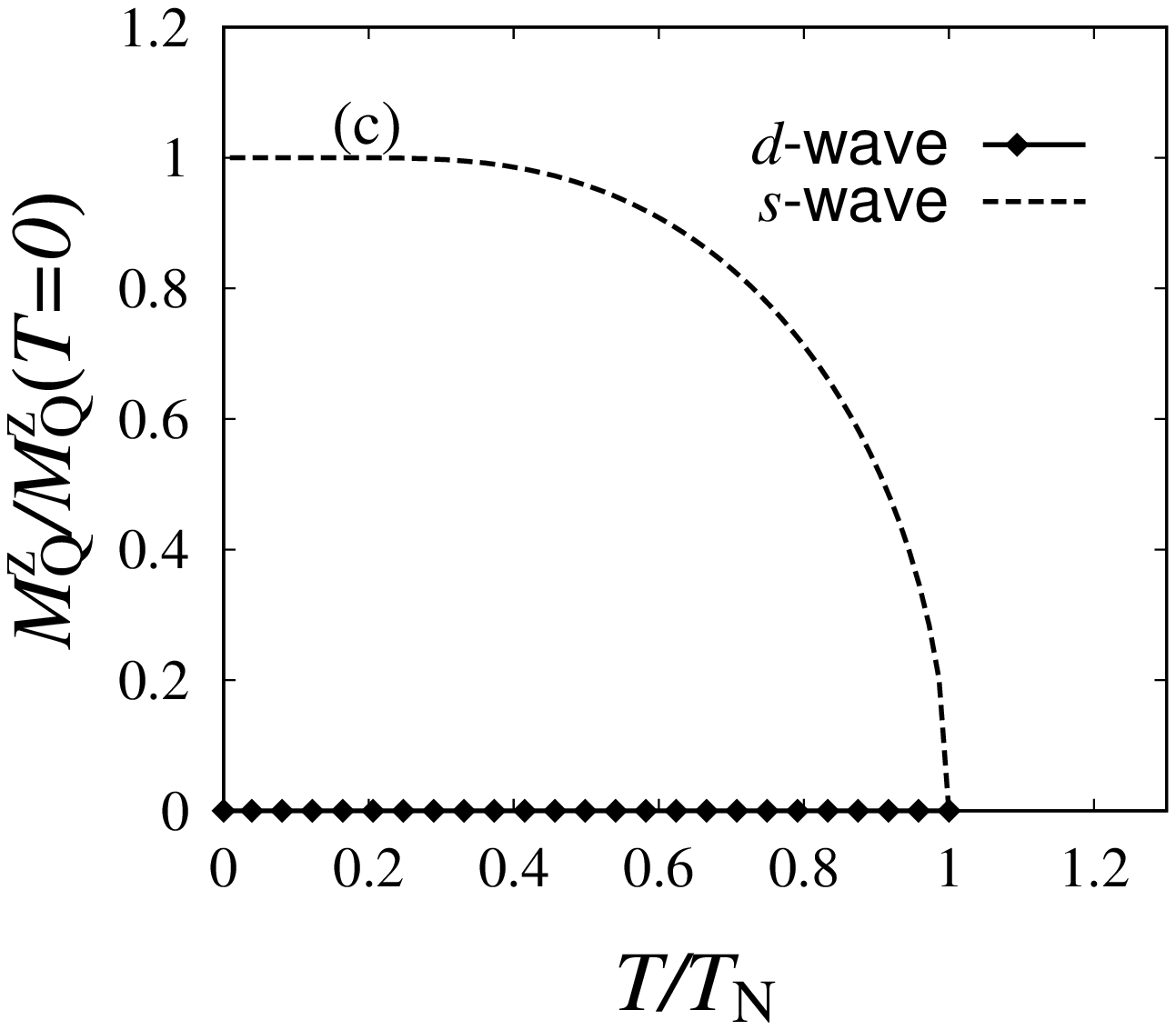}
\end{center}
\caption{Temperature dependences of (a) the specific heat, (b) the uniform
susceptibility, and (c) the staggered magnetic moment. These are respectively
scaled to $C(T=T_\mathrm{N}+0)$, $\chi_0(T=T_\mathrm{N})$ and $M^z_Q(T=0)$.
In each figure, the temperature is normalized by $T_\mathrm{N}$ of each state;
the discrepancy in $T>T_\mathrm{N}$ arises from the difference of
$T_\mathrm{N}$ in between the $s$-wave and the $d$-wave. In Fig.(b) $\chi_{zz}$
and $\chi_{xx}$ are respectively $\chi_{0\parallel}$ and $\chi_{0\perp}$ in
this text.}
\label{fig:3}
\end{figure}

\noindent
(A) Figure~\ref{fig:3}(a): Specific heat $C$.
\begin{equation}
\label{eq:C}
C(T)=\sum_k
  \left( \dfrac{E_k^2}{T}-\dfrac{1}{2}\diff{|\Delta^{d,s}_k|^2}{T} \right)
  \diff{}{E_k}\th\dfrac{E_k}{2T}.
\end{equation}
The specific heat shows a discontinuous jump at $T_\mathrm{N}$ as in the case
of the conventional SDW. The origin of the jump is the
$d|\Delta^{d,s}_k|^2/dT$-term in Eq.~(\ref{eq:C}) because of
$|\Delta^{d,s}_k|\propto\sqrt{T_\mathrm{N}-T}$ near $T_\mathrm{N}$.
\par
\noindent
(B) Figure~\ref{fig:3}(b): Uniform susceptibility $\chi_0$.
\begin{equation}
\left\{
\begin{array}{l}
\dis \chi_{0\parallel} = \dfrac{1}{2}\sum_k \diff{}{E_k} \th\dfrac{E_k}{2T},\\
\dis \chi_{0\perp} = \dfrac{1}{2}\sum_k
   \left(  \dfrac{|\Delta^{d,s}_k|^2}{E_k^3}\th\dfrac{E_k}{2T}
          +\dfrac{\xi_k^2}{E_k^2}\diff{}{E_k} \th\dfrac{E_k}{2T}
   \right),
\end{array}
\right.
\end{equation}
where $\chi_{0\parallel}$ ($\chi_{0\perp}$) is the parallel (perpendicular)
component to the z-axis. $\chi_{0\perp}$ is almost constant below
$T_\mathrm{N}$, while $\chi_{0\parallel}$ shows a rapid decrease due to the
reduction of the density of states below the energy gap; their behaviors are
just equal to those in the usual antiferromagnetism.
\par
Since the density of states behaves like $N(E)\propto E~(E\sim 0)$ in the case
of the $d$-wave SDW, $C$ and $\chi_0$ show power-law temperature dependences at
$T\ll T_\mathrm{N}$ as $C\propto T^2$ and $\chi_{0\parallel}\propto T$. On the
other hand, they show exponential decreases in the case of the conventional
$s$-wave SDW due to the finite excitation gap.
\par
{\it (3) Magnetic moment:}
The staggered magnetic moment is given by
\begin{equation}
\label{eq:M}
M^z_Q=\sum_k \left<\sigma^z_{kk+Q}\right>
       =\sum_k\dfrac{\Delta^{d,s}_k}{2E_k}\th\dfrac{E_k}{2T}.
\end{equation}
In the case of $d$-wave SDW, Eq.~(\ref{eq:M}) always vanishes because of
$\Delta^d_k\propto\phi_k^d$. Namely, the spin density itself is homogeneous in
spite of the name, ``spin density wave''\cite{rf:comment1}. We emphasize that
the absence of $M^z_Q$ is qualitatively different from the case of the
conventional $s$-wave SDW in which $M^z_Q$ is essentially equivalent to the
order parameter. We show their qualitative difference in Fig.~\ref{fig:3}(c).
\par
The absence of $M^z_Q$ means that the antiferromagnetic susceptibility does not
diverges at $T_\mathrm{N}$ in contrast to the conventional $s$-wave SDW. In the
latter case, the divergence leads to a deviation of NMR-$T_1^{-1}$ from the
Korringa-like temperature dependence.
On the other hand, the $T_1^{-1}\propto T$ is expected up to just above
$T_\mathrm{N}$ in the $d$-wave SDW because of the absence of the divergence in
the spin susceptibility. This result agrees with the experiment\cite{rf:Tou}
\par
How does the ``micro-'' but ``finite-'' staggered moment appear in the present
$d$-wave SDW? We have two answers for this question: (1) When the lattice is
deformed to some extent, the cancellation of the summation in Eq.~(\ref{eq:M})
becomes incomplete because of the deformation of the Fermi surface.
Then $M^z_Q$ can be finite. In this case, the staggered magnetic moment may
strongly depend on the applied pressure; such a behavior is actually reported
in UPt$_3$\cite{rf:Hayden}.
(2) Since the unconventional SDW should be destroyed near impurities and
boundaries, the $s$-wave SDW may be induced around them by the proximity
effect, as in the case of the superconductivity. In this case, we obtain
a finite staggered magnetic moment with a short range correlation, because
the staggered magnetic moment which accompanys the $s$-wave component localizes
around the defects. We explain details of these mechanisms in our forthcoming
paper\cite{rf:comment2}.
\par
%% Conclusion %%
{\it Summary and Discussion}---
We have proposed a novel SDW characterized by the electron-hole pair amplitude
changing its sign in the momentum space as a possible mechanism of the
micromagnetism in URu$_2$Si$_2$. This kind of unconventional SDW shows clear
anomalies in the specific heat and the uniform susceptibility as in the case of
the usual SDW, while the induced staggered magnetic moment is absent.
We emphasize that these results well explain the properties of the
micromagnetism in URu$_2$Si$_2$. Although we examined the $d$-wave SDW in two
dimension as a simple example, the essential results in this paper describe
general property of the unconventional SDW.
\par
Finally, we briefly comment on UPt$_3$ which has no anomaly in, for instance,
the specific heat. When we consider the order parameter which is zero at the
Fermi surface, such as an extended $s$-wave symmetry or an odd-$\omega$ state
discussed in the field of superconductivity, the jump in the specific heat
becomes small or absent. Thus, we can expect that the unconventional SDW is
a possible mechanism also for the micromagnetism in this material.
\par
%% acknowledgments %%
{\it Acknowledgements}---
We would like to give thanks to Professor S. Takada for useful discussions and
reading this manuscript. We would like to thank Professors K. Terakura,
M. Sigrist and D. Hirashima for valuable comments. One of the author (H.I.) is
also grateful to Dr. H. Tou, Professors Y. Kohori, M. Imada, K. Miyake,
Dr. Y. Onishi and colleagues in JRCAT for their interesting comments.
This work was partly supported by NEDO.
%% references %%

%% Figure captions %%

%--- Fig.1 ---

%--- Fig.2 ---

%--- Fig.3 ---

%% Figures %%

%----- end document -------

\begin{thebibliography}{99}
\bibitem{rf:Aeppli}
G. Aeppli, E. Bucher, C. Broholm, J. K. Kjems, J. Baumann and J. Hfnagl,
Phys. Rev. Lett. {\bf 60}, 615 (1988).
\bibitem{rf:Isaacs1}
E. D. Isaacs, P. Zschack, C. L. Broholm, C. Burns, G. Aeppli, A. P. Ramirez,
T. T. M. Palstra, R. W. Erwin, N. St\"ucheli and E. Bucher,
Phys. Rev. Lett. {\bf 75}, 1178 (1995).
\bibitem{rf:Broholm}
C. Broholm, H. Lin, P. T. Matthews, T. E. Mason, W. J. L. Buyers,
M. F. Collins, A. A. Menovsky, J. A. Mydosh, and J. K. Kjems,
Phys. Rev {\bf B43}, 12809 (1991).
\bibitem{rf:Isaacs2}
E. D. Isaacs, D. B. McWhan, R. N. Kleiman, D. J. Bishop, G. E. Ice, P. Zschack,
B. D. Gauin, T. E. Mason, J. D. Garrett, and W. J. L. Buyers,
Phys. Rev. Lett. {\bf 65}, 3185 (1990).
\bibitem{rf:Visser}
A. de Visser, A. Menovsky and J. J. M. Franse,
Physica {\bf B147}, 81 (1987).
\bibitem{rf:Palstra}
T. T. M. Palstra, A. A. Menovsky, J. van den Berg, A. J. Dirkmaat, P. H. Kes,
G. J. Nieuwenhuys, and J. A. Mydosh,
Phys. Rev. Lett. {\bf 55}, 2727 (1985).
\bibitem{rf:Tou}
H. Tou, Y. Kitaoka, K. Asayama, N. Kimura, Y. \=Onuki, E. Yamamoto and
K. Maezawa,
Phys. Rev. Lett. {\bf 77}, 1374 (1996).
\bibitem{rf:Santini}
P. Santini and G. Amoretti,
Phys. Rev. Lett. {\bf 73}, 1027 (1994).
\bibitem{rf:Barzykin}
V. Barzykin and L. P. Gor'kov,
Phys. Rev. Lett. {\bf 74}, 4301 (1995),
$ibid.$ {\bf 70}, 2479 (1993).
\bibitem{rf:Sikkema}
A. E. Sikkema, W. J. L. Buyers, I. Affleck and J. Gan,
Phys. Rev. {\bf B54}, 9322 (1996).
\bibitem{rf:Kimura}
N. Kimura, R. Settai, Y. \=Onuki, H. Toshima, E. Yamamoto, K. Maezawa, H. Aoki
and H. Harima,
J. Phys. Soc. Jpn {\bf 64}, 3881 (1995).
\bibitem{rf:Maple}
M. B. Maple, J. W. Chen, Y. Dalichaouch, T. Kohara, C. Rossel,
M. S. Torikachvili, M. W. McElfresh and J. D. Thompson,
Phys. Rev. Lett. {\bf 56}, 185 (1986).
\bibitem{rf:Ozaki}
M. Ozaki,
Int. J. Quantum. Chem. {\bf 42}, 55 (1992).
\bibitem{rf:comment1}
As a problem of terminology, Ozaki call this state the spin-current-wave-state,
because it has a finite spin current. In this paper, however, we still use the
word ``SDW'' in spite of the absent of the spin density wave in order to
emphasize the extension of the conventional SDW.
\bibitem{rf:Hayden}
S. M. Hayden, L. Taillefer, C. Vettier and J. Flouquet,
Phys. Rev. {\bf B46},  8675 (1992).
\bibitem{rf:comment2}
H. Ikeda and Y. Ohashi,
in preparation.
\end{thebibliography}
\end{document}